\documentclass[11pt]{article}
\usepackage{graphics}
\usepackage{cite}
\begin{document}
\def\up{\uparrow}
\def\dn{\downarrow}
\def\Up{\Uparrow}
\def\Dn{\Downarrow}
\def\ket#1{|#1\rangle}
\def\bra#1{\langle#1|}
\def\eff{{\rm eff}}
\def\Vec#1{\mbox{\boldmath $#1$}}
\Large

\centerline{\bf Inversion Phenomena of the Anisotropies}
\centerline{\bf of the Hamiltonian and the Wave-Function}
\centerline{\bf in the Distorted Diamond Type Spin Chain}

\bigskip
\normalsize
\centerline{\Large Kiyomi Okamoto${}^*$ and Yutaka Ichikawa}

\bigskip

\centerline{\Large\it Department of Physics, Tokyo Institute of Technology,}
\medskip

\centerline{\Large\it Oh-okayama, Meguro-ku, Tokyo 152-8551, Japan}

\bigskip
\bigskip

We investigate the ground-sate phase diagram of the $XXZ$ version of the $S=1/2$
distorted diamond chain by use of the degenerate perturbation theory
near the truncation point.
In case of the $XY$-like interaction anisotropy, the phase diagram consists of
the N\'eel phase and the spin-fluid phase.
For the Ising-like interaction anisotropy case, it consists of three phases:
the ferrimagnetic phase, the N\'eel phase and the spin-fluid phase.
The magnetization in the ferrimagnetic phase is $1/3$ of the saturation
magnetization.
The remarkable nature of the phase diagram is the existence of the 
N\'eel phase,
although the interaction anisotropy is $XY$-like.
And also, the spin-fluid phase appears in spite of the
Ising-like interaction anisotropy.
We call these regions \lq\lq inversion regions\rq\rq.
\bigskip
\bigskip
\noindent

\bigskip
\noindent
*Corresponding author: \\
E-mail: kokamoto@stat.phys.titech.ac.jp \\
Fax:+81-3-5734-2739

\newpage
%
%
%
%
%
%
\bigskip
\bigskip

Very recently the $S=1/2$ distorted diamond (DD) chain was
studied both theoretically \cite{okamoto,tone1,tone2,tone3,sano,honecker}
and experimentally \cite{ishii,kikuchi}.
The illustration of the DD chain is shown in figure 1.
In this letter we study the ground-state of the $XXZ$ version of
the DD chain near the truncation point 
$J_2 = J_3 =0$ by the degenerate perturbation theory
and show the inversion phenomena between the anisotropies of 
the Hamiltonian and the wave-function.
The Hamiltonian of the $XXZ$ DD chain is expressed by
\begin{eqnarray}
   H
   =&&J_1 \sum_j \left\{h_{3j-1,3j}(\Delta) + h_{3j,3j+1}(\Delta) \right\}
    + J_2 \sum_j h_{3j+1,3j+2}(\Delta) \nonumber \\
    &&+ J_3 \sum_j \left\{ h_{3j-2,3j}(\Delta) + h_{3j,3j+2}(\Delta) \right\}
   \label{eq:ham}
\end{eqnarray}
where
\begin{equation}
    h_{l,m}(\Delta)
    = S_l^x S_m^x + S_l^y S_m^y + \Delta S_l^z S_m^z 
\end{equation}
and $\Delta>0$ denotes the $XXZ$ anisotropy.
All the coupling constants are supposed to be positive (antiferromagnetic).
The intra-trimer coupling $J_1$ is supposed to be larger than 
the inter-trimer couplings $J_2$ and $J_3$.

\begin{figure}[h]
      \begin{center}
         \scalebox{0.5}[0.5]{\includegraphics{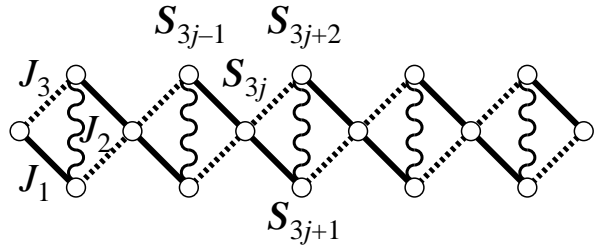}}
      \caption{Illustration of the distorted diamond (DD) chain.
      The number of the trimer is denoted by $j$.
      Solid lines represent the intra-trimer coupling $J_1$,
      wavy lines the inter-trimer coupling $J_2$,
      and dotted lines the inter-trimer coupling $J_3$.}
      \end{center}
\end{figure}
Let us consider the $J_2 = J_3 =0$ case,
where the problem is reduced to the 3-spin problem.
The ground states of a trimer are
\begin{eqnarray}
    \ket{\psi_j^{(+)}}
    &=& {1 \over \sqrt{a^2+2}}
      \left( \ket{\up\up\dn}
             - a\ket{\up\dn\up}
             + \ket{\dn\up\up} \right)
      ~~~~~(S_{\rm tot}^z = +1) \\
    \ket{\psi_j^{(-)}}
    &=& {1 \over \sqrt{a^2+2}}
      \left( \ket{\dn\dn\up} 
             - a\ket{\dn\up\dn}
             + \ket{\up\dn\dn} \right)
      ~~~~~(S_{\rm tot}^z = -1) 
\end{eqnarray}
where $\ket{\up\up\dn}$ is the abbreviation of 
$\ket{\up_{3j-1}\up_{3j}\dn_{3j+1}}$, and
\begin{equation}
    a = {\Delta + \sqrt{\Delta^2 + 8} \over 2}
\end{equation}
The ground-state energy $E_0$ per a trimer is
\begin{equation}
    {E_0 \over J_1}
    = - {\Delta^2 + \sqrt{\Delta^2 + 8} \over 4}
\end{equation}
It is easy to see
\begin{eqnarray}
    &&\bra{\psi^+} S_{3j-1}^z \ket{\psi^+} 
    = \bra{\psi^+} S_{3j+1}^z \ket{\psi^+}
    = {a^2 \over 2(a^2+2)} > 0 \\
    &&\bra{\psi^+} S_{3j}^z \ket{\psi^+}
    = - {a^2-2 \over 2(a^2+2)} < 0 \\
    &&\bra{\psi^-} S_{3j-1}^z \ket{\psi^-} 
    = \bra{\psi^-} S_{3j+1}^z \ket{\psi^-}
    = -{a^2 \over 2(a^2+2)} < 0 \\
    &&\bra{\psi^-} S_{3j}^z \ket{\psi^-}
    = {a^2-2 \over 2(a^2+2)} > 0
\end{eqnarray}
because $a>\sqrt{2}$ for $\Delta>0$.
Near the truncation point $J_2 = J_3 =0$,
we can restrict ourselves to these two states (equation (3) and (4)) for a trimer,
neglecting other 6 states.
For convenience we represent these two states by a pseudo-spin $\Vec{T}_j$
\begin{equation}
    \ket{\Up_j} = \ket{\psi_j^{(+)}},~~~~~
    \ket{\Dn_j} = \ket{\psi_j^{(-)}}
\end{equation}
where $\ket{\Up_j}$ and $\ket{\Dn_j}$ denote
the $T_j^z = 1/2$  and $T_j^z = 1/2$ states, respectively.
The interactions between trimers are expressed as the interactions
between pseudo-spins.
For instance, we have
\begin{equation}
    {J_3 \over 2}
    \left( S_{3j}^+ S_{3j+2}^- + S_{3j}^- S_{3j+2}^+ \right)
    \Rightarrow
    - {2aJ_3 \over (a^2+2)^2}
    \left(T_j^+ T_{j+1}^- + T_j^- T_{j+1}^+\right)
\end{equation}
A straightforward calculation leads to
\begin{equation}
    H_{\rm eff}
    = \sum_j \left\{
        J_{\rm eff}^\perp \left(T_j^x T_{j+1}^x + T_j^y T_{j+1}^y \right)
        + J_{\rm eff}^z T_j^z T_{j+1}^z \right\}
\end{equation}
where
\begin{eqnarray}
    &&J_{\rm eff}^\perp
    = {a \over (a^2+2)^2}\left\{ 4aJ_2 - 8J_3\right\} \\
    &&J_{\rm eff}^z
    = {a \over (a^2+2)^2}\left\{ a^2(a^2-2)J_2 - 2(a^2-2)^2J_3) \right\}
\end{eqnarray}
When $\Delta=1$, we see $J_{\rm eff}^\perp = J_{\rm eff}^z = (4/9)(J_2-J_3)$.
Then the ground state of $H_{\rm eff}$ is either the ferromagnetic state
or the spin-fluid state depending on whether $J_2<J_3$ or $J_2>J_3$.
The ferromagnetic state of $\Vec{T}$ corresponds to
the ferrimagnetic state of $\Vec{S}$ with the magnetization of $M_{\rm s}/3$,
where $M_{\rm s}$ is the saturation magnetization.
This explains the boundary between the spin-fluid and ferrimagnetic states
of the phase diagram of the isotropic Heisenberg DD chain \cite{tone1,honecker}.
The phase diagram for the $\Delta=1$ case is shown in figure 2,
where $\tilde J_2=J_2/J_1$ and $\tilde J_3=J_3/J_1$.

\begin{figure}[h]
      \begin{center}
         \scalebox{0.38}[0.38]{\includegraphics{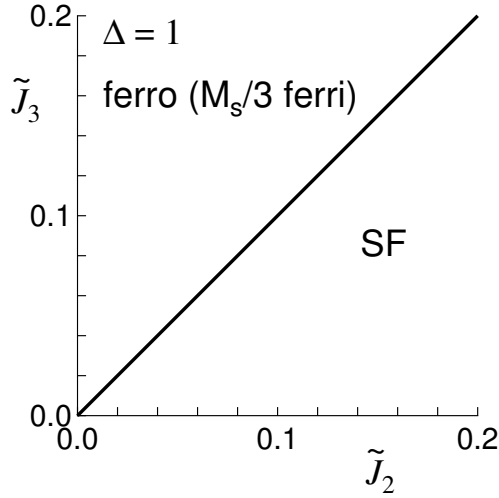}}
      \caption{Phase diagrams for the $\Delta=1$ case.
      SF denotes the spin-fluid state.
      The ferromagnetic state in the $\Vec{T}$-picture corresponds to the
      $M_{\rm s}/3$ ferrimagnetic state in the $\Vec{S}$-picture. 
      The slope of the boundary line is unity.}
      \end{center}
\end{figure}
In general case ($\Delta \ne 1$),
the ground state of $H_{\rm eff}$ is known from $J_\eff^\perp$ and $J_\eff^z$
as 
\begin{equation}
  \matrix{
    J_\eff^z <0 \hbox{ and } |J_\eff^z| > |J_\eff^\perp|
    &~~\Rightarrow~~
    &\hbox{ferromagnetic state} \hfill\cr
    |J_\eff^z| < |J_\eff^\perp| \hfill
    &~~\Rightarrow~~
    &\hbox{spin-fluid state} \hfill\cr
    J_\eff^z > |J_\eff^\perp| \hfill
    &~~\Rightarrow~~
    &\hbox{N\'eel state} \hfill
  }
  \label{eq:condition}
\end{equation}
When $\Delta <1$ ($XY$-like case),
the ferromagnetic state does not appear.
From the condition (\ref{eq:condition}), 
we see
\begin{equation}
  \matrix{
    J_3 < c_1 J_2 \hfill     &~~\Rightarrow~ &\hbox{spin-fluid state}   \cr
    c_1 J_2 < J_3 < c_2 J_2  &~~\Rightarrow~ &\hbox{N\'eel state}\hfill \cr
    J_3 > c_2 J_2 \hfill     &~~\Rightarrow~ &\hbox{spin-fluid state}   }
\end{equation}
where
\begin{equation}
    c_1 = {a^2+2a+2 \over 2a(a+2)}~~~~~~~
    c_2 = {1 \over 2}{a(a+2)(a^2-2a+2) \over a^4-4a^2+8}
\end{equation}
The phase diagram for the $\Delta=0.5$ case 
near the truncation point $J_2 = J_3 =0$ is shown
in figure 3.
We note that 
\begin{equation}
    c_1 \to {5 \over 8}~~~~~c_2 \to 1~~~~~(\hbox{as }\Delta \to 1)
\end{equation}

\begin{figure}[h]
      \begin{center}
         \scalebox{0.38}[0.38]{\includegraphics{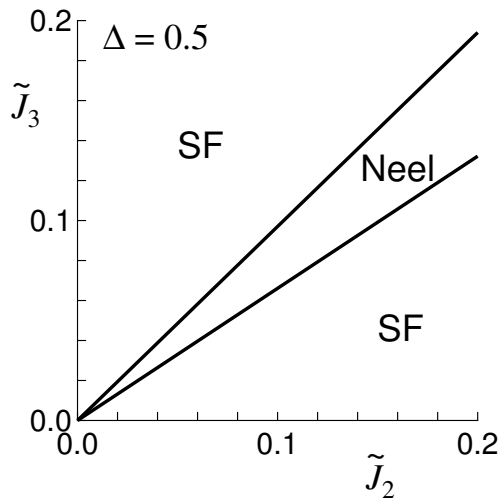}}
      \end{center}
      \caption{Phase diagrams for the $\Delta=0.5$ ($XY$-like) case.
      The slopes of the boundary lines are $c_1=0.6609$ (lower line)
      and $c_2 = 0.9703$ (upper line), respectively.}
   \hskip0.5truecm
\end{figure}
For the  $\Delta >1$ (Ising-like) case, we have
\begin{equation}
  \matrix{
    J_3 < c_1 J_2 \hfill     &~~\Rightarrow~ &\hbox{N\'eel state}\hfill   \cr
    c_1 J_2 < J_3 < c_2 J_2  &~~\Rightarrow~ &\hbox{spin-fluid state}\hfill \cr
    J_3 > c_2 J_2 \hfill     &~~\Rightarrow~ &\hbox{ferromagnetic state}\hfill   }
\end{equation}
In the Ising limit $\Delta \to \infty$,
we see $c_1 = c_2 = 1/2$, which results in the vanishment of the
spin-fluid state in figure 4.
The striking nature of figure 4 is the existence of the
N\'eel region,
although the interaction anisotropy is $XY$-like. 
Similarly, the spin-fluid state is realized in figure 4
in spite of the Ising-like anisotropy.
These are the {\it inversion phases}.

\begin{figure}[h]
      \begin{center}
         \scalebox{0.38}[0.38]{\includegraphics{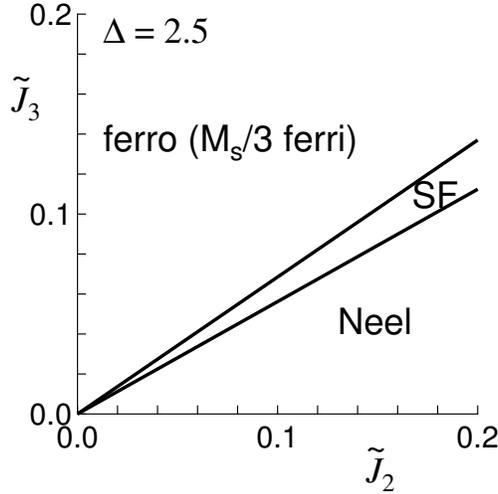}}
      \end{center}
      \caption{Phase diagrams for the $\Delta = 2.5$ case.
      The slopes of the boundary lines $c_1=0.5620$ (lower line),
      and $c_2 = 0.6850$ (upper line), respectively.}
\end{figure}
Let us consider the physical reason for these {\it inversion phenomena}.
The present model has three characteristics:
the frustrations, the trimer nature and the $XXZ$ anisotropy.
In case  of strong frustrations, 
the spin system is going to form singlet pairs
to avoid the energy loss by competing interactions \cite{MG,Majumdar,ON}.
But the formation of singlet pairs is incompatible with the trimer nature.
Thus, in the $XY$-like interaction case,
the spin system turn to the $z$-direction to avoid the energy loss, 
because interaction of the $z$-direction is weaker than that of the
$xy$-direction in the $XY$-like case.
Similarly, we can explain the existence of the spin-fluid state
for the Ising-like interaction case.
This {\it inversion phenomena} is due to the interplay
among the frustration, the trimer nature and the $XXZ$ anisotropy of the
Hamiltonian.

When $\Delta=1$, the ferrimagnetic state in figure 2 is highly degenerate.
The states $S_{\rm tot}^z = M_{\rm s}/3, M_{\rm s}/3-1,\cdots, -M_{\rm s}/3$
have the same energy due to the $SU(2)$ symmetry.
When the interaction anisotropy is introduced,
this degeneracy is lifted.
The lowest-energy state is with $S_{\rm tot}^z =0$ in the $XY$-like case,
whereas with $S_{\rm tot}^z = 0$ in the Ising-like case.
Thus the behavior of the upper left region of the phase diagram
is physically explained.
The lower right region of the phase diagram 
is essentially the same as the uniform chain
with nearest neighbor interactions,
which is realized at the point $\tilde J_3=0, \tilde J_2=1$.
When $\tilde J_3$ is sufficiently large, the dimer state is expected
as in case of $\Delta=1$ \cite{okamoto}.
However, such a region is beyond the description of the present degenerate
perturbation theory.

In conclusion, we have investigated the phase diagram of the
$S=1/2$ distorted diamond chain with $XXZ$ interaction anisotropy.
We have found the {\it inversion phenomena} between
the Hamiltonian anisotropy and the wave-function anisotropy.
As far as we know, this is the first finding of the {\it inversion phenomena}.
Our preliminary numerical calculations support the present phase diagrams.
The full phase diagrams of the present model for a wider range of
parameters will be published elsewhere.

\section*{Acknowledgement}

We wish to thank Takashi Tonegawa and Makoto Kaburagi
for fruitful discussions.

\end{document}